\documentstyle[aps,preprint,tighten,eqsecnum,floats,epsf,rotate,prd]{revtex}

\newcommand{\del}{\partial}

\newcommand{\rf}[4]{{\em {#1}} {\bf #2}, #3 (#4)}

\newcommand{\pr}{Phys.\ Rev.\ }

\newcommand{\physl}{Phys.\ Lett.\ }

\newcommand{\np}{Nucl.\ Phys.\ }
\newcommand{\beq}{\begin{equation}}
\newcommand{\eeq}{\end{equation}}
\newcommand{\beqa}{\begin{eqnarray}}
\newcommand{\eeqa}{\end{eqnarray}}

\newcommand{\muhat}{\hat{\mu}}

\newbox\rotbox

\begin{document}

\preprint{\vbox{
\rightline{ADP-99-21/T363}
\rightline{JLAB-THY-99-10}} \\
}

\title{Discretisation Errors in Landau Gauge on the Lattice}

\author{Frederic D.\ R.\ Bonnet\footnote{
E-mail:~fbonnet@physics.adelaide.edu.au},
Patrick O.\ Bowman\footnote{
E-mail:~pbowman@physics.adelaide.edu.au \hfill\break
\null\quad\quad
WWW:~http://www.physics.adelaide.edu.au/$\sim$pbowman/}
and Derek B.\ Leinweber\footnote{
E-mail:~dleinweb@physics.adelaide.edu.au ~$\bullet$~ Tel:
+61~8~8303~3548 ~$\bullet$~ Fax: +61~8~8303~3551 \hfill\break
\null\quad\quad
WWW:~http://www.physics.adelaide.edu.au/theory/staff/leinweber/}}
\address{Special Research Centre for the Subatomic Structure of Matter
and The Department of Physics and Mathematical Physics, University of
Adelaide, Adelaide SA 5005, Australia}
\author{Anthony G.\ Williams\footnote{
E-mail:~awilliam@physics.adelaide.edu.au ~$\bullet$~ Tel:
+61~8~8303~3546 ~$\bullet$~ Fax: +61~8~8303~3551 \hfill\break
\null\quad\quad
WWW:~http://www.physics.adelaide.edu.au/theory/staff/williams.html}}
\address{Special Research Centre for the Subatomic Structure of Matter
and The Department of Physics and Mathematical Physics, University of
Adelaide, Adelaide SA 5005, Australia \\ and Department of Physics and SCRI, 
Florida State University, Tallahassee, Florida, USA}
\author{David G.\ Richards\footnote{
E-mail:~dgr@jlab.org \hfill\break
\null\quad\quad}}

\address{Dept. of Physics and Astronomy, University of Edinburgh, Edinburgh 
EH9 3JZ, Scotland \\ and
Thomas Jefferson National Accelerator Facility, 12000 Jefferson Avenue,
Newport News, VA 23606, USA.}

\date{\today}

\maketitle

\begin{abstract} 
Lattice discretisation errors in the Landau gauge condition are
examined.  An improved gauge fixing algorithm in which ${\cal O}(a^2)$
errors are removed is presented.  ${\cal O}(a^2)$ improvement of the gauge 
fixing condition improves comparison with continuum Landau gauge in two ways:
1) through the elimination of ${\cal O}(a^2)$ errors and 2) through a
secondary effect of reducing the size of higher-order errors.
These results emphasise the importance of implementing an improved gauge
fixing condition.
\end{abstract}


\section{Introduction}
\label{sec:intro}

Gauge fixing in lattice gauge theory simulations is crucial for many
calculations.  It is required for the study of gauge dependent
quantities such as the gluon propagator\cite{rapid_glu}, and is used to 
facilitate other techniques such as gauge-dependent fermion source smearing.
However, the standard Landau
gauge condition\cite{cthd} is the same as the continuum condition,
$\sum_\mu \del_\mu A_\mu = 0$, only to leading order in the lattice
spacing $a$.  

There has been some study of alternate lattice definitions of the
Landau gauge condition \cite{lg}.  The focus of this letter is to use
mean-field-improved perturbation theory\cite{tadpole} to compare  
different lattice definitions of the Landau gauge, and quantify the sizes of
the discretisation errors. In particular, we derive a new
${\cal O}(a^2)$ improved Landau-gauge-fixing functional which is central to
estimating the discretisation errors made with the standard
functionals.

In section~\ref{sec:formalism} we derive mean-field-improved
expansions (in the lattice spacing and coupling constant) for three
different definitions of the Landau gauge condition for the lattice.
The use of an improved Landau gauge functional allows an estimate of the
absolute error in standard lattice Landau gauge.  The size of the
error provides a strong argument for the use of an improved gauge
fixing condition.

\section{Lattice Landau Gauge}
\label{sec:formalism}

Gauge fixing on the lattice is achieved by maximising a functional whose
extremum implies the gauge fixing condition.  The usual Landau 
gauge fixing functional is \cite{cthd}
\begin{equation}
{\cal F}^{G}_{1}[\{U\}] = \sum_{\mu, x}\frac{1}{2} \mbox{Tr} \, \left\{
U^{G}_{\mu}(x) + U^{G}_{\mu}(x)^{\dagger} \right\},
\label{eqn:f1}
\end{equation}
where
\begin{equation}
U^{G}_{\mu}(x) = G(x) U_{\mu}(x) G(x+\hat{\mu})^{\dagger},
\end{equation}
and
\begin{equation}
G(x) = \exp \left\{ -i \sum_a \omega^a(x) T^a \right\}.
\end{equation}
Taking the functional derivative of eqn~(\ref{eqn:f1}), we obtain
\begin{equation}
\frac{\delta {\cal F}^{G}_{1}}{\delta \omega^a(x)} = \frac{1}{2} i 
\sum_{\mu} \mbox{Tr} \, \left\{ \left [ U^{G}_{\mu}(x-\hat{\mu}) 
- U^{G}_{\mu}(x) - \left ( U^{G}_{\mu}(x-\hat{\mu}) - U^{G}_{\mu}(x) 
\right )^{\dagger} \right ] T^{a} \right\}. \label{eqn:deltaf1}
\end{equation}
The gauge links are defined as
\beq
U_\mu(x) \equiv {\cal P}\exp \left\{ i g \int_0^a dt A_\mu(x+\muhat t) 
\right\}.
\eeq
Connection with the continuum is made by Taylor-expanding $A_\mu(x+\muhat t)$
 about $x$, integrating term-by-term, and then expanding the exponential, 
typically to leading order in $g$, noting
that errors are of ${\cal O}(g^2 a^2)$.  Expanding
eqn~(\ref{eqn:deltaf1}), we obtain
\begin{equation}
\frac{\delta {\cal F}^{G}_{1}}{\delta \omega^a(x)} = g a^2 \sum_{\mu}
\mbox{Tr} \left\{ \left [ \partial_{\mu} A_{\mu}(x) + \frac{1}{12} a^2
\del_\mu^3 A_{\mu}(x) + \frac{a^4}{360} \del_\mu^5 A_\mu(x) 
+ {\cal O}(a^6) \right ] T^a \right\} + {\cal O}(g^3 a^4).
\label{eqn:deltaf1_expand}
\end{equation}

To lowest order in $a$, an extremum of 
eqn~(\ref{eqn:f1}) satisfies $\sum_\mu \del_\mu A_\mu(x) = 0$, which is 
the continuum Landau gauge condition.  What it means in practice on the lattice
is that
\beq
\sum_\mu \del_\mu A_\mu(x) 
= \sum_\mu \left\{ -\frac{a^2}{12} \del_\mu^3 A_\mu(x) 
- {\cal H}_1 \right\},
\label{eqn:err1}
\eeq
where ${\cal H}_1$ represents ${\cal O}(a^4)$ and higher-order terms,
as shown in  
eqn~(\ref{eqn:deltaf1_expand}).
Na\"{\i}vely one might hope that higher-order derivatives in the
brackets of (\ref{eqn:deltaf1_expand}) are small, but it will be shown
that the terms on the R.H.S. of eqn~(\ref{eqn:err1}) are
large compared to the numerical accuracy possible in gauge fixing
algorithms.

An alternative gauge-fixing functional can be constructed using two-link 
terms, for example
\begin{equation}
{\cal F}^{G}_2 = \sum_{x,\mu} \frac{1}{2} \mbox{Tr} \left\{
U^{G}_{\mu}(x) U^{G}_{\mu}(x+\hat{\mu}) + \mbox{h.c.} \right\}.
\label{eqn:f2}
\end{equation}
Taking the functional derivative yields
\begin{equation}
\frac{\delta {\cal F}^{G}_{2}}{\delta \omega^a(x)} = \frac{1}{2} i
\sum_{\mu} \mbox{Tr} \left\{ \left [ U^{G}_{\mu}(x-2\hat{\mu})
U^{G}_{\mu}(x-\hat{\mu}) - U^{G}_{\mu}(x) U^{G}_{\mu}(x+\hat{\mu}) -
\mbox{h.c.} \right ] T^a \right\}, \label{eqn:deltaf2}
\end{equation}
and expanding to ${\cal O}(a^4)$ we obtain
\begin{equation}
\frac{\delta {\cal F}^{G}_{2}}{\delta \omega^a(x)} = 4 g a^2 \sum_{\mu}
\mbox{Tr} \left\{ \left [ \del_{\mu} A_{\mu}(x) + \frac{a^2}{3} 
\partial_{\mu}^3 A_{\mu}(x) + \frac{16}{360} a^4 \del_\mu^5 A_\mu(x)
	+ {\cal O}(a^6) \right ] T^a \right\} + {\cal O}(g^3 a^4), 
\label{eqn:deltaf2_expand}
\end{equation}
which again implies the continuum
Landau-gauge-fixing condition to lowest order in $a$.

${\cal O}(a^2)$ errors can be removed from the gauge fixing condition
by taking a linear combination of the one-link and two-link
functionals:
\begin{equation}
\frac{\delta \left\{ \frac{4}{3} {\cal F}^{G}_1 - \frac{1}{12 u_0} {\cal
F}^{G}_2 \right\} }{\delta \omega^a(x)} = g a^2 \sum_{\mu} \mbox{Tr}
\left\{ \left [ \del_{\mu} A_{\mu}(x) - \frac{4}{360} a^4 \del_\mu^5 
A_\mu(x) + {\cal O}(a^6) \right ] T^a \right\} + {\cal O}(g^3 a^4)
\label{eqn:f_improved}
\end{equation}
where we have introduced the mean-field (tadpole) improvement parameter 
$u_0$ to ensure that our perturbative calculation is not spoiled by large
renormalisations~\cite{tadpole}.  For the tadpole-improvement parameter we 
employ the plaquette measure
\beq
u_0=\left(\frac{1}{3}{\cal R}eTr<U_{\text{pl}}>\right)^{\frac{1}{4}}.
\eeq
and note that the higher order $g^3 a^4$ terms of 
eqns~(\ref{eqn:deltaf1_expand})~(\ref{eqn:deltaf2_expand})~and
~(\ref{eqn:f_improved}) are to be viewed in terms of the mean-field-improved 
perturbation theory~\cite{tadpole}.  For future reference we shall define
the improved functional, ${\cal F}_{\text{Imp}}^G \equiv \frac{4}{3}
{\cal F}_1^G - \frac{1}{12 u_0} {\cal F}_2^G$.

Once a gauge-fixing functional has been defined,
an algorithm must be chosen to perform the gauge fixing.  We adopt a 
``steepest descents'' approach \cite{cthd}.  Collecting terms of 
${\cal O}(a^4)$ and higher into the ${\cal H}_i$, we define
\beqa
\Delta_1(x)  & \equiv & \frac{1}{u_0} \sum_\mu \left [ U_\mu(x-\mu) 
		- U_\mu(x) - \mbox{h.c.} 
 		\right ]_{\text{traceless}} \nonumber \\
	     & = & -2 i g a^2 \sum_\mu \left\{ \del_{\mu} A_{\mu}(x) 
		+ \frac{a^2}{12} \del_{\mu}^3 A_{\mu}(x)  
                + {\cal H}_1 \right\} \ ,
\label{eqn:delta1_expand}
\eeqa

\beqa
\Delta_2(x)  & \equiv & \frac{1}{4 u_0^2} \sum_\mu \left [ 
		U_\mu(x-2\mu)U_\mu(x-\mu) - U_\mu(x)U_\mu(x+\mu) 
		- \mbox{h.c.} \right ]_{\text{traceless}} \nonumber \\
	     & = & -2 i g a^2 \sum_\mu \left\{ \del_{\mu} A_{\mu}(x) 
		+ \frac{a^2}{3} \del_{\mu}^3 A_{\mu}(x)
                + {\cal H}_2 \right\} \ ,
\label{eqn:delta2_expand}
\eeqa

\beqa
\Delta_{\text{Imp}}(x) & \equiv & \frac{4}{3} \Delta_1(x) 
				- \frac{1}{3} \Delta_2(x) \nonumber \\
	& = & -2 i g a^2 \sum_\mu \left\{ \del_{\mu} A_{\mu}(x)
                + {\cal H}_{\text{Imp}} \right\}.
\label{eqn:DImp_expand}
\eeqa
where the subscript, ``traceless'' denotes subtraction of the average of the
colour-trace from each of the diagonal colour elements.
The resulting gauge transformation is
\beqa
G_i(x) & = & \exp \left\{ \frac{\alpha}{2} \Delta_i(x) \right\} \nonumber \\
       & = & 1 + \frac{\alpha}{2} \Delta_i(x) + {\cal O}(\alpha^2),
\label{eqn:gauge_trans}
\eeqa
where $\alpha$ is a tuneable step-size parameter, and the index $i$ is either 
$1$, $2$, or $\text{Imp}$.  At each iteration $G_i(x)$ is unitarised through
an orthonormalisation procedure.
The gauge fixing algorithm proceeds by calculating
the relevant $\Delta_i$ in terms of the mean-field-improved links, and then
applying the associated gauge transformation, eqn~(\ref{eqn:gauge_trans}),
to the gauge field.  The algorithm is implemented in parallel, updating all 
links simultaneously, and is iterated until the Lattice Landau gauge condition
is satisfied, to within some numerical accuracy.

\section{Discretisation Errors in the Gauge Fixing Condition}
\label{sec:analysis}

The approach to Landau gauge is usually measured by a quantity such as
\beq
\theta_i = \frac{1}{V N_c} \sum_x \mbox{Tr} \left\{ \Delta_i(x) 
\Delta_i(x)^{\dagger}
\right\}
\eeq
which should tend to zero as the configuration becomes gauge fixed.  $N_c$ is
the number of colours, i.e. $3$.

A configuration fixed using $\Delta_1(x)$ will satisfy eqn~(\ref{eqn:err1}).
Substituting eqn~(\ref{eqn:err1}) into eqn~(\ref{eqn:delta2_expand}) yields
\beqa
\Delta_2(x) & = & -2iga^2 \sum_\mu \left\{ 
		-\frac{a^2}{12} \del_{\mu}^3 A_{\mu}(x) 
		+ \frac{a^2}{3} \del_{\mu}^3 A_{\mu}(x) 
		- {\cal H}_1 + {\cal H}_2 \right\} \nonumber \\
	    & = & -2iga^2 \sum_\mu \left\{ 
		\frac{a^2}{4} \del_{\mu}^3 A_{\mu}(x) 
		- {\cal H}_1 + {\cal H}_2 \right\}
\eeqa
and similarly,
\beq
\Delta_{\text{\text{Imp}}}(x) = -2iga^2 \sum_\mu \left\{ -\frac{a^2}{12} 
	\del_{\mu}^3 A_{\mu}(x) - {\cal H}_1 + {\cal H}_{\text{Imp}} \right\}.
\label{eqn:deltaImp_err}
\eeq
Since the improved measure has no ${\cal O}(a^2)$ error of its own,
eqn~(\ref{eqn:deltaImp_err}) provides an estimate of the absolute size of 
these discretisation errors.

\section{Calculations on the Lattice}
\label{sec:results}

\subsection{The Gauge Action.}

The ${\cal O}(a^2)$ tadpole-improved action is defined as
\beq
S_G=\frac{5\beta}{3}\sum_{\text{pl}}{\cal R}eTr(1-U_{\text{pl}}(x)) 
	- \frac{\beta}{12u_{0}^2}\sum_{\text{rect}}
	{\cal R}eTr(1-U_{\text{rect}}(x)),
\label{gaugeaction}
\eeq
where the operators $U_{\text{pl}}(x)$ and $U_{\text{rect}}(x)$ are defined 
\beqa
U_{\text{pl}}(x) & = & U_{\mu}(x)U_{\nu}(x+\hat{\mu})
		U^{\dagger}_{\mu}(x+\hat{\nu}) U^\dagger_{\nu}(x),
\eeqa
and
\beqa
U_{\text{rect}}(x) & = & U_{\mu}(x)U_{\nu}(x+\hat{\mu})
		U_{\nu}(x+\hat{\nu}+\hat{\mu})
		U^{\dagger}_{\mu}(x+2\hat{\nu})U^{\dagger}_{\nu}(x+\hat{\nu})
		U^\dagger_{\nu}(x) \nonumber \\
	& + & U_{\mu}(x)U_{\mu}(x+\hat{\mu})U_{\nu}(x+2\hat{\mu})
		U^{\dagger}_{\mu}(x+\hat{\mu}+\hat{\nu})
		U^{\dagger}_{\mu}(x+\hat{\nu})U^\dagger_{\nu}(x).
\eeqa
The link product $U_{\text{rect}}(x)$ denotes the rectangular
$1\times2$ and $2\times1$ plaquettes.  Eqn~(\ref{gaugeaction})
reproduces the continuum action as $a\rightarrow{0}$, provided that
$\beta$ takes the standard value of $6/g^2$. ${\cal O}(g^2a^2)$ corrections
to this action are estimated to be of the order of two to three
percent~\cite{Alf95}.  Note that our $\beta=6/g^2$ differs from that
used in~\cite{Alf95,Lee,Woloshyn}.  Multiplication of our $\beta$ in
eqn~(\ref{gaugeaction}) by a factor of $5/3$ reproduces their
definition.

\subsection{Numerical Simulations}

Gauge configurations are generated using the
Cabbibo-Marinari~\cite{Cab82} pseudoheat-bath algorithm with three
diagonal $SU(2)$ subgroups.  The mean link, $u_0$, is averaged every
10 sweeps and updated during thermalisation.  

For the exploration of gauge fixing errors we consider
$6^4$ lattices at a variety of $\beta$ for both standard Wilson and
improved actions.  For the standard Wilson action we consider $\beta =
5.7$, 6.0, and 6.2, corresponding to lattice spacings of 0.18, 0.10,
and 0.07 fm respectively.  For the improved action we consider $\beta
= 3.92$, 4.38, and 5.00, corresponding to lattice spacings of
approximately 0.3, 0.2, and 0.1 fm respectively.

\begin{figure}[t]
\begin{center}
\epsfysize=13truecm
\leavevmode
\setbox\rotbox=\vbox{\epsfbox{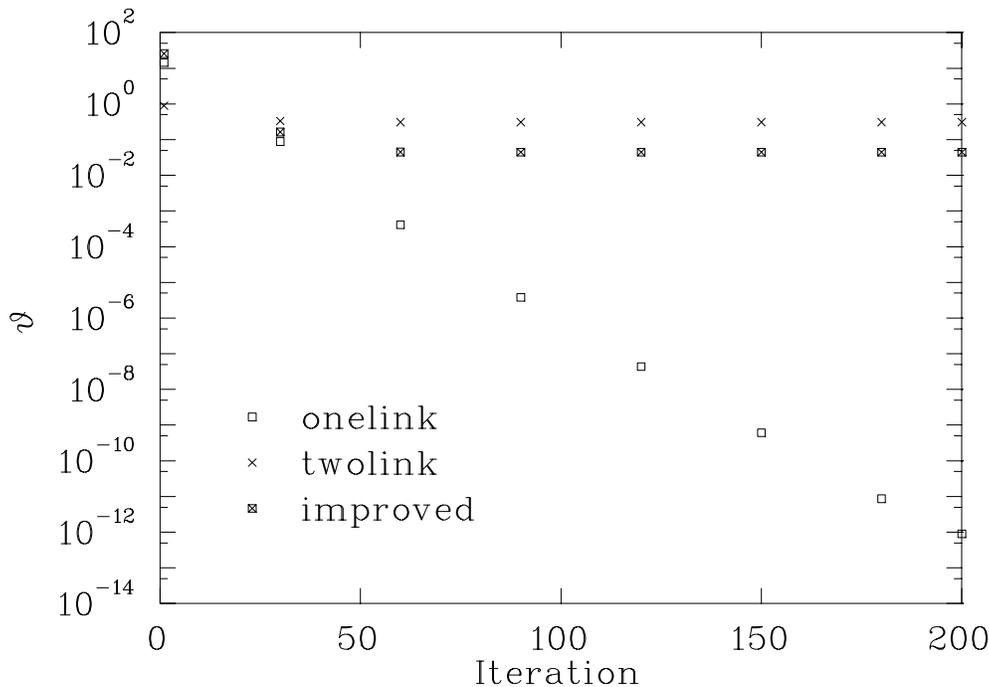}}\rotl\rotbox
\end{center}
\caption{The gauge fixing measures for a $6^4$ lattice with Wilson action 
at $\beta = 6.0$.  This lattice was gauge fixed with $\Delta_1$, so 
$\theta_1$ drops steadily whilst $\theta_2$ and $\theta_{\text{Imp}}$
plateau at much higher values.}
\label{fig:thetas}
\end{figure}

The configurations were gauge fixed, using Conjugate Gradient Fourier 
Acceleration \cite{cm} until $\theta_1 < 10^{-12}$.  $\theta_{\text{Imp}}$
and $\theta_2$ were then measured, to see the size of the residual higher
order terms.  The evolution of the gauge fixing measures is shown for one of
the lattices in fig.~\ref{fig:thetas}.  This procedure was then repeated, 
fixing with each of the other two functionals, and
the results are shown in table~\ref{tab:wil_s6t6}.  The same procedure
was performed with three ${\cal O}(a^2)$-improved lattices, and the results 
are shown in table~\ref{tab:imp_s6t6}.

\begin{table}
\begin{tabular}{ddclll}
$\beta$ & $u_0$ & Functional & $\theta_{\text{Imp}}$ & $\theta_2$ 
& $\frac{\theta_{\text{Imp}}}{\theta_2}$ \\ 
\hline
  5.7   & 0.865 & 1 & 0.0679 & 0.611 & 0.111 \\
  6.0   & 0.877 & 1 & 0.0578 & 0.520 & 0.111 \\
  6.2   & 0.886 & 1 & 0.0522 & 0.469 & 0.111 \\
\hline
\hline
 \\
\hline
\hline
$\beta$ & $u_0$ & Functional & $\theta_{\text{Imp}}$ & $\theta_1$
& $\frac{\theta_1}{\theta_{\text{Imp}}}$ \\ 
\hline
  5.7   & 0.865 & 2 &  59.0  &  33.2 & 0.563 \\
  6.0   & 0.877 & 2 &  65.1  &  36.6 & 0.563 \\
  6.2   & 0.886 & 2 &  61.7  &  34.7 & 0.563 \\
\hline
\hline
 \\
\hline
\hline
$\beta$ & $u_0$ & Functional &$\theta_1$ & $\theta_2$ 
& $\frac{\theta_1}{\theta_2}$ \\ 
\hline
  5.7   & 0.865 & Imp & 0.0427 & 0.684 & 0.0625 \\
  6.0   & 0.877 & Imp & 0.0367 & 0.588 & 0.0625 \\
  6.2   & 0.886 & Imp & 0.0332 & 0.531 & 0.0625  
\end{tabular}
\caption{Values of the gauge-fixing measures obtained using the Wilson gluon 
action on $6^4$ lattices at three values of the lattice spacing, fixed to
Landau gauge with the one-link, two-link and improved functionals 
respectively.}
\label{tab:wil_s6t6}
\end{table}

\begin{table}
\begin{tabular}{ddclll}
$\beta$ & $u_0$ & Functional & $\theta_{\text{Imp}}$ & $\theta_2$
& $\frac{\theta_{\text{Imp}}}{\theta_2}$ \\ \hline
  3.92  & 0.837 & 1 & 0.102  & 0.921 & 0.111 \\
  4.38  & 0.880 & 1 & 0.0585 & 0.526 & 0.111 \\
  5.00  & 0.904 & 1 & 0.0410 & 0.369 & 0.111 \\
\hline
\hline
 \\
\hline
\hline
$\beta$ & $u_0$ & Functional & $\theta_{\text{Imp}}$ & $\theta_1$ 
& $\frac{\theta_1}{\theta_{\text{Imp}}}$ \\ 
\hline
  3.92  & 0.837 & 2 &  57.5  & 32.3 & 0.563 \\
  4.38  & 0.880 & 2 &  53.4  & 30.0 & 0.563 \\
  5.00  & 0.904 & 2 &  52.2  & 29.4 & 0.563 \\
\hline
\hline
 \\
\hline
\hline
$\beta$ & $u_0$ & Functional & $\theta_1$ & $\theta_2$ 
& $\frac{\theta_1}{\theta_2}$ \\ 
\hline
  3.92  & 0.837 & Imp & 0.0638 & 1.02  & 0.0625 \\
  4.38  & 0.880 & Imp & 0.0366 & 0.586 & 0.0625 \\
  5.00  & 0.904 & Imp & 0.0261 & 0.417 & 0.0625
\end{tabular}
\caption{Values of the gauge-fixing measures obtained using the improved 
gluon action on $6^4$ lattices at three values of the lattice spacing, fixed to
Landau gauge with the one-link, two-link and improved functionals 
respectively.}
\label{tab:imp_s6t6}
\end{table}

Comparing eqn~(\ref{eqn:err1}) with eqn~(\ref{eqn:deltaImp_err}), we can see
that if we fix a configuration to Landau gauge by using the basic, one-link
functional, the improved measure will consist entirely of the 
discretisation errors.  Looking at table~\ref{tab:wil_s6t6}, we see that at
$\beta = 6.0$, $\theta_{\text{Imp}} = 0.058$, i.e.
\beq
\frac{1}{V N_c} \sum_{x} \mbox{Tr} \left \{\sum_\mu \del_\mu A_\mu(x)
\left (\sum_\nu \del_\nu A_\nu(x) \right )^{\dagger} \right \} 
= 0.058,
\eeq
a substantial deviation from the continuum Landau gauge when compared with
$\theta_1 < 10^{-12}$.

As a check of our simulations, we note that the definition for 
$\Delta_{\text{Imp}}$, eqn~(\ref{eqn:DImp_expand}), provides a constraint
on the measures when gauge fixed.  For example
\beq
\frac{\theta_{\text{Imp}}}{\theta_2}  =  \frac{(-\frac{1}{12})^2}
		{(-\frac{1}{12} + \frac{1}{3})^2}
		=  \frac{1}{9} \simeq  0.111.
\eeq
Similarly, by fixing with $\Delta_2(x)$ we expect
\beq
\frac{\theta_1}{\theta_{\text{Imp}}} =  \frac{(-\frac{1}{3} 
		+ \frac{1}{12})^2} {(-\frac{1}{3})^2} 
		 = \frac{9}{16} \simeq  0.563,
\eeq
and fixing with $\Delta_{\text{Imp}}(x)$ leads to
\beq
\frac{\theta_1}{\theta_2} = \frac{1}{16} = 0.0625.
\eeq
These ratios are reproduced by the data of tables \ref{tab:wil_s6t6} and
\ref{tab:imp_s6t6}.

A three-link functional can also be constructed, e.g.
\beq
{\cal F}_3^G [\{ U \}] = \sum_{\mu, x} \frac{1}{2} \mbox{Tr} 
	\left\{U_\mu^G(x-\mu) U_\mu^G(x) U_\mu^G(x+\mu) + \mbox{h.c.} \right\},
\eeq
the functional derivative of which,
\begin{eqnarray}
\frac{\delta {\cal F}_3^G}{\delta \omega^a(x)} &=&
	\frac{1}{2} i \sum_\mu \mbox{Tr} \left\{ \left [ 
	U_\mu^G(x-3\mu) U_\mu^G(x-2\mu) U_\mu^G(x-\mu) 
        \right . \right . \nonumber \\
        &&\hspace{2cm} \left . \left .
	- U_\mu^G(x) U_\mu^G(x+\mu) U_\mu^G(x+2\mu) - \mbox{h.c.} \right ] 
	T^a \right\}
\end{eqnarray}
leads to
\beq
\Delta_3(x) = -2iga^2 \sum_\mu \left\{ \del_\mu A_\mu(x) 
	+ \frac{3}{4} a^2 \del_\mu^3 A_\mu(x) 
	+ \frac{9}{40} a^4 \del_\mu^5 A_\mu(x) + {\cal O}(a^6) \right\}
	+ {\cal O}(g^3 a^4).
\eeq
If leading order errors dominate, then we should be able to make ratios
like those above, but involving $\theta_3$.  However, we find that the values
taken from our simulations are very different from ratios based solely on 
leading, ${\cal O}(a^2)$ errors.  This indicates that the sum of higher-order 
errors is also significant.
Whilst one could proceed to combine $\Delta_1$, $\Delta_2$ and 
$\Delta_{\text{Imp}}$ to eliminate both ${\cal O}(a^2)$ and ${\cal O}(a^4)$,
it is likely that ${\cal O}(g^2a^2)$ errors are of similar size to the 
${\cal O}(a^4)$ errors.  We will defer such an investigation to future work.

A configuration fixed using $\Delta_{\text{Imp}}(x)$ will satisfy 
\beq
\sum_\mu \del_\mu A_\mu(x) 
= \sum_\mu \left\{ - {\cal H}_{\text{Imp}} \right\}.
\label{eqn:errImp}
\eeq
Substituting this into eqn~(\ref{eqn:delta1_expand}) yields
\beq
\Delta_1(x) = -2iga^2 \sum_\mu \left\{ 
		\frac{a^2}{12} \del_\mu^3 A_\mu(x)  
		+ {\cal H}_1 - {\cal H}_{\text{Imp}} \right\}.
\eeq
A comparison of this equation with eqn~(\ref{eqn:deltaImp_err}) reveals that
the coefficients of the terms in curly brackets, expressing the discretisation
errors in these two cases, differ only by an overall
sign, which is lost in the calculation of the corresponding $\theta_i$.
If the three different methods presented all fixed in exactly the same way,
then the $\theta_{\text{Imp}}$ of a configuration fixed with $\Delta_1$, would
be equal to $\theta_1$ when the configuration is fixed with 
$\Delta_{\text{Imp}}$.  It is clear from the tables that they are not, 
signaling the higher-order derivative terms $\del_\mu^n A_\mu(x)$ take
different values depending on the gauge fixing functional used.

Examining the values in tables \ref{tab:wil_s6t6} and
\ref{tab:imp_s6t6} reveals that
in every case $\theta_1$ is smaller when we have fixed with the improved
functional than $\theta_{\text{Imp}}$ under the one-link functional.  This
suggests that the additional long range information used by the improved 
functional is producing a gauge fixed configuration with smaller, higher-order
derivatives; a secondary effect of improvement.

Equally, one can compare the value of $\theta_2$ when fixed using the
one-link functional, and $\theta_1$ when fixed using the two-link
functional.  In this case, their differences are rather large and are
once again attributed to differences in the size of higher-order
derivatives of the gauge field.  The two-link functional is
coarser, knows little about short range fluctuations, and fails to
constrain higher-order derivatives.  Similar conclusions are
drawn from a comparision of $\theta_2$ fixed with the improved
functional and $\theta_{\text{Imp}}$ fixed with the two link
functional.

It is also interesting to note that in terms of the absolute errors,
the Wilson action at $\beta = 6.0$ is comparable to the improved
lattice at $\beta = 4.38$, where the lattice spacing is three times
larger.

\section{Conclusions}

We have fixed gluon field configurations to Landau gauge by three
different functionals: one-link and a two-link functionals, both with
${\cal O}(a^2)$ errors, and an improved functional, with ${\cal
O}(a^4)$ errors.  Using these functionals we have devised a method for
estimating the discretisation errors involved.  Our results indicate
that order ${\cal O}(a^2)$ improvement of the gauge fixing condition
will improve comparison with the continuum Landau gauge in two ways:
1) through the elimination of ${\cal O}(a^2)$ errors and 2) through a
secondary effect of reducing the size of higher-order errors.  These
conclusions are robust with respect to lattice spacing and we have
also verified the stability of our conclusions by considering
additional configurations to that presented here.  We plan to
investigate improved gauge fixing on larger volume lattices to see if
these effects of improvement persist.  Lattice Landau gauge, in its
standard implementation, is substantially different from its continuum
counterpart, despite fixing the Lattice gauge condition to one part in
$10^{12}$.

\section*{Acknowledgements}

Thanks to Francis Vaughan and the Cooperative Research Centre for
Advanced Computational Systems (ACSys) for support in the development
of parallel algorithms implemented in Connection Machine Fortran (CMF).
DGR thanks Mike Peardon for helpful conversations in the early stages
of this work.  This research is supported by the Australian Research
Council.  DGR acknowledges the support of PPARC through an Advanced
Fellowship, and thanks FNAL for their hospitality during the course of
this work.


\end{document}